\documentclass{article}
\usepackage{amsfonts}

\usepackage{graphicx}
\usepackage{amsmath}

\input{tcilatex}

\begin{document}

\title{From bi-Hamiltonian geometry to separation of variables:
stationary Harry-Dym and the KdV dressing chain}
\author{Maciej B\l aszak \footnote{Supported partially by KBN research grant
No. 5 P03B 004 20}\\
%EndAName
Institute of Physics, A.Mickiewicz University,\\
Umultowska 85, 61-614 Pozna\'{n}, Poland}

%\Adress{Institute of Physics, A. Mickiewicz University, \\
%Umultowska 85, 61-614 Pozna\'{n}, Poland}

\maketitle

\begin{abstract}
Separability theory of one-Casimir Poisson pencils, written down
in arbitrary coordinates, is presented. Separation of variables
for stationary Harry-Dym and the KdV dressing chain illustrates
the theory.
\end{abstract}

\section{Introduction}

The separation of variables is one of the most important methods of solving
nonlinear ordinary differential equations of Hamiltonian type. It is known
since 19$th$ century, when Hamilton and Jacobi proved that given a set of
appropriate coordinates, the so called separated coordinates, it is possible
to solve a related Liouville integrable dynamical system by quadratures.
Unfortunately in the $19th$ century and most of the $20th$ century, for a
number of models of classical mechanics the separated variables were either
guessed or found by some \emph{ad hoc }methods. A fundamental progress in
this field was made in 1985, when Sklyanin adopted the method of soliton
systems, i.e. the Lax representation, to systematic construction of
separated variables (see his review article \cite{sk1}). In his approach,
the appropriate Hamiltonians appear as coefficients of the spectral curve,
i.e. the characteristic equation of the Lax matrix. Recently, a new
constructive separability theory was presented, based on a bi-Hamiltonian
property of integrable systems. In the frame of canonical coordinates the
theory was developed in a series of papers \cite{1}-\cite{6} (see also the
review article \cite{7}), while the general case was considered in \cite{m1}
and \cite{m2}.

In this paper we briefly summarize the results of the theory in the case of
one-Casimir Poisson pencils and illustrate it on two examples: the
stationary flow of Harry-Dym (canonical coordinates frame) and the KdV
dressing chain (noncanonical coordinates frame). This last system is
separated for the first time. Finally, on the basis of examples, we make a
few comments on the relation between a separation curve of the
bi-Hamiltonian approach and a spectral curve of the Sklyanin approach.

\section{Preliminary considerations}

Let $M$ be a differentiable manifold, $TM$ and $T^{\ast }M$ its tangent and
cotangent bundle. At any point $u\in M,$ the tangent and cotangent spaces
are denoted by $T_{u}M$ and $T_{u}^{\ast }M$, respectively. The pairing
between them is given by the map $<\cdot ,\cdot >:$ $T_{u}^{\ast }M\times
T_{u}M\rightarrow \mathbb{R}.$ For each smooth function $F\in C^{\infty }(M),
$ $dF$ denotes the differential of $F$. $M$ is said to be a Poisson manifold
if it is endowed with a Poisson bracket $\{\cdot ,\cdot \}:C^{\infty
}(M)\times $ $C^{\infty }(M)\rightarrow C^{\infty }(M),$ in general
degenerate. The related Poisson tensor $\pi $ is defined by $\{F,G\}_{\pi
}(u)\,:=<dG,\pi \circ dF>(u)\,=<dG(u),\pi (u)dF(u)>$. So, at each point $%
u,\,\pi (u)$ is a linear map $\pi (u):T_{u}^{\ast }M\rightarrow T_{u}M$
which is skew-symmetric and has vanishing Schouten bracket with itself, i.e.
the related bracket fulfils the Jacobi identity. Any function $c\in
C^{\infty }(M),$ such that $dc\in \ker \pi ,$ is called a Casimir of $\pi .$
Let $\pi _{0}$,$\pi _{1}:T^{\ast }M\rightarrow TM$ be two Poisson tensors on
$M.$ A vector field $K$ is said to be  bi-Hamiltonian with respect to $\pi
_{0}$ and $\pi _{1}$ if there exist two smooth functions $H$,$F\in C^{\infty
}(M)$ such that
\begin{equation}
K=\pi _{0}\circ dH=\pi _{1}\circ dF.  \label{0}
\end{equation}%
The Poisson tensors $\pi _{0}$ and $\pi _{1}$ are said to be compatible if
the associated pencil $\pi _{\lambda }=\pi _{1}-\lambda \pi _{0}$ is itself
a Poisson tensor for any $\lambda \in \mathbb{R}.$

In this paper we consider a particular Poisson manifold $M$ of $\dim M=2n+1$
equipped with a linear Poisson pencil $\pi _{\lambda }$ of maximal rank.
Assuming that a Casimir of the pencil is a polynomial in $\lambda $ of an
order $n$
\begin{equation}
h_{\lambda }=h_{0}\lambda ^{n}+h_{1}\lambda ^{n-1}+...+h_{n}  \label{2}
\end{equation}%
one gets a bi-Hamiltonian chain
\begin{equation}
\pi _{\lambda }\circ dh_{\lambda }=0\Longleftrightarrow
\begin{array}{l}
\pi _{0}\circ dh_{0}=0 \\
\pi _{0}\circ dh_{1}=K_{1}=\pi _{1}\circ dh_{0} \\
\pi _{0}\circ dh_{2}=K_{2}=\pi _{1}\circ dh_{1} \\
\,\,\,\,\,\,\,\,\,\,\,\,\,\,\,\,\,\,\,\,\,\,\,\,\,\,\,\,\,\,\,\,\vdots  \\
\pi _{0}\circ dh_{n}=K_{n}=\pi _{1}\circ dh_{n-1} \\
\qquad \qquad \,\,\,\,\,\,\,\,\,\,\,0=\pi _{1}\circ dh_{n}.\,\,%
\end{array}
\label{3}
\end{equation}%
where $\{h_{i}\}_{i=1}^{n}$ is a set of independent functions in involution
with respect to both Poisson structures, so defines a Liouville integrable
system on $M$.

\emph{When is the system separated?} Let us introduce a set of canonical
coordinates $\{\lambda _{i},\mu _{i}\}_{i=1}^{n}$ and a Casimir coordinate $%
c=h_{0}.$ Then, let us linearize the system through a canonical
transformation $(\mu ,\lambda )\rightarrow (a,b)$ in the form $b_{i}=\frac{%
\partial W}{\partial a_{i}},\mu _{i}=\frac{\partial W}{\partial \lambda _{i}}%
,$ where $W(\lambda ,a)$ is a generating function satisfying the
related Hamilton-Jacobi (HJ) equations
\begin{equation}
h_{r}(\lambda ,\frac{\partial W}{\partial \lambda })=a_{r},\;\;\;\;%
\;r=1,...,n.  \label{5}
\end{equation}%
In general, HJ equations (\ref{5}) are nonlinear partial
differential equations and to solve them is a hoopless task.
Nevertheless, one can find a complete integral in some special
case, when in $(\mu ,\lambda )$ coordinates a generating function
$W$ is additively separated:
\begin{equation}
W(\lambda ,a)=\sum_{i=1}^{n}W_{i}(\lambda _{i},a).  \label{6}
\end{equation}%
In such a case HJ equations turn into a set of decoupled ordinary
differential equations and hence, at least in principle, can be solved by
quadratures. Then, in $(a,b)$ coordinates the flow is trivial
\begin{equation}
(a_{j})_{t_{r}}=0,\,\,\,\,(b_{j})_{t_{r}}=\delta _{jr}  \label{8}
\end{equation}%
and the implicit form of the trajectories $\lambda _{i}(t_{r})$ is
\begin{equation}
b_{j}(\lambda ,a)=\frac{\partial W}{\partial a_{j}}=\delta
_{jr}t_{r}+const,\,\,\,\,j=1,...,n.  \label{9}
\end{equation}%
Such $(\lambda ,\mu )$ coordinates are called \emph{separated coordinates.}

\textbf{Lemma 1} \emph{A sufficient condition for $(\lambda ,\mu )$ to be
separated coordinates for the bi-Hamiltonian chain (\ref{3}) is
\begin{equation}
h_{\lambda _{i}}=f_{i}(\lambda _{i},\mu _{i}),\;\;\;i=1,...,n,  \label{11}
\end{equation}%
where
\[
h_{\lambda _{i}}=c\lambda _{i}^{n}+h_{1}\lambda _{i}^{n-1}+...+h_{n}
\]%
and $f_{i}(\lambda _{i},\mu _{i})$ is an arbitrary smooth function of a pair
of canonically conjugate coordinates.}

\textbf{Proof} Using the following notation
\begin{equation}
h=(c,h_{1},...,h_{n})^{T},\;\;\;v_{i}=(\lambda _{i}^{n},\lambda
_{i}^{n-1},...,\lambda _{i}^{0}=1),  \nonumber
\end{equation}%
\begin{equation}
v=(v_{1},...,v_{n})^{T},\;\;\;f=(f_{1},...,f_{n})^{T},  \label{13}
\end{equation}%
the condition (\ref{11}) can be presented in the matrix form
\begin{equation}
v\cdot h=f\Longleftrightarrow \left(
\begin{array}{cccc}
\lambda _{1}^{n} & \lambda _{1}^{n-1} & \cdots  & 1 \\
\lambda _{2}^{n} & \lambda _{2}^{n-1} & \cdots  & 1 \\
\vdots  & \vdots  & \cdots  & \vdots  \\
\lambda _{n}^{n} & \lambda _{n}^{n-1} & \cdots  & 1%
\end{array}%
\right) \left(
\begin{array}{c}
c \\
h_{1} \\
\vdots  \\
h_{n}%
\end{array}%
\right) =\left(
\begin{array}{c}
f_{1}(\lambda _{1},\mu _{1}) \\
f_{2}(\lambda _{2},\mu _{2}) \\
\vdots  \\
f_{n}(\lambda _{n},\mu _{n})%
\end{array}%
\right)   \label{14}
\end{equation}%
which may be called a \emph{generalized St\"{a}ckel representation. }Indeed,
\begin{equation}
f_{i}=(v\cdot h)_{i}=v_{i}\cdot h=h_{\lambda _{i}}.  \label{15}
\end{equation}%
Multiplying the HJ equations (\ref{5}), written in the matrix form
\begin{equation}
h=a,\;\;\;a=(c,a_{1},...,a_{n})^{T},  \label{16}
\end{equation}%
from the left by $v_{i}$ one gets
\begin{equation}
v_{i}\cdot h=v_{i}\cdot a\Longrightarrow f_{i}(\lambda _{i},\frac{\partial W%
}{\partial \lambda _{i}})=c\lambda _{i}^{n}+a_{1}\lambda
_{i}^{n-1}+...+a_{n}\Longrightarrow W(\lambda
,a)=\sum_{i=1}^{n}W_{i}(\lambda _{i},a).  \label{17}
\end{equation}

\section{Separated coordinates}

In ref. \cite{1} the bi-Hamiltonian chain (\ref{3}) in the separated
coordinates $(\lambda ,\mu ,c)$ was constructed for the first time.
Actually, the Hamiltonian functions $h_{k}$ take the following compact form
\begin{equation}
h_{k}(\lambda ,\mu ,c)=-\sum_{i=1}^{n}\frac{\partial \rho _{k}(\lambda )}{%
\partial \lambda _{i}}\frac{f_{i}(\lambda _{i},\mu _{i})}{\Delta
_{i}(\lambda )}+c\rho _{k}(\lambda ),\,\,\,k=1,...,n  \label{18}
\end{equation}%
where $\,\,\Delta _{i}(\lambda ):=\prod_{j\neq i}(\lambda _{i}-\lambda _{j})$%
, $\rho _{k}(\lambda )$ are the elementary symmetric polynomials and the two
Poisson structures are
\begin{equation}
\pi _{0}=\left(
\begin{array}{rrr}
0 & I & 0 \\
-I & 0 & 0 \\
0 & 0 & 0%
\end{array}%
\right) ,\,\,\,\pi _{1}=\left(
\begin{array}{rrr}
0\,\,\,\,\,\,\,\,\,\, & \Lambda \,\,\,\,\,\,\,\,\,\,\, & h_{1,\mu } \\
-\Lambda \,\,\,\,\,\,\,\,\,\, & 0\,\,\,\,\,\,\,\,\,\,\,\, & -h_{1,\lambda }
\\
-\left( h_{1,\mu }\right) ^{T} & \left( h_{1,\lambda }\right) ^{T} &
0\,\,\,\,%
\end{array}%
\right) ,  \label{20}
\end{equation}%
where $\Lambda =diag(\lambda _{1},...,\lambda _{n})$ and $h_{1,\mu }:=\left(
\frac{\partial h_{1}}{\partial \mu _{1}},...,\frac{\partial h_{1}}{\partial
\mu _{n}}\right) ^{T}$. Notice that all $h_{k}$ are linear in $c$. In fact
there exists a family of separated coordinates $(\lambda ^{\prime },\mu
^{\prime },c)$ which preserve the form (\ref{18}) and (\ref{20}), and are
related to the set $(\lambda ,\mu ,c)$ by a canonical transformation
\begin{equation}
\lambda _{i}^{\prime }=\lambda _{i},\;\;\;\mu _{i}^{\prime }=\mu
_{i}+\vartheta _{i}(\lambda _{i}),\;\;\;i=1,...,n,  \label{19}
\end{equation}%
where $\vartheta _{i}$ are arbitrary smooth function.

If $f_{i}=f,\;i=1,...,n,$ then the separated coordinates are $n$ different
points of a curve
\begin{equation}
f(\lambda ,\mu )=h_{\lambda },\,\,\,\,\,\,\,\,\,\,\,\,\,\,\,\,\,h_{\lambda
}=c\lambda ^{n}+h_{1}\lambda ^{n-1}+...+h_{n},  \label{21}
\end{equation}%
called the \emph{separation curve. }

In the separated coordinates a Poisson pencil and the chain can be trivially
projected onto a symplectic leaf $S$ of $\pi _{0}$ ($\dim S=2n$) as  $\theta
_{\lambda }=\theta _{1}-\lambda \theta _{0},$ where
\begin{equation}
\theta _{0}=\left(
\begin{array}{cc}
0 & I \\
-I & 0%
\end{array}%
\right) ,\,\,\,\,\,\theta _{1}=\left(
\begin{array}{cc}
0 & \Lambda  \\
-\Lambda  & 0%
\end{array}%
\right) ,  \label{25}
\end{equation}%
is a nondegenerate Poisson pencil on $S$. Hence, $S$ is Poisson-Nijenhuis
manifold where the related Nijenhuis tensor $N$
\begin{equation}
N=\theta _{1}\circ \theta _{0}^{-1}=%
\begin{array}{cc}
\Lambda  & 0 \\
0 & \Lambda
\end{array}
\label{26}
\end{equation}%
and its adjoint $N^{\ast }$ are diagonal. This is the reason why $(\lambda
,\mu )$ are called the Darboux-Nijenhuis (DN) coordinates. On $S$ the chain (%
\ref{3}),(\ref{18}),(\ref{20}) gives rise to
\begin{equation}
N^{\ast }\circ d\widehat{h}_{i}=d\widehat{h}_{i+1}-\rho _{i}d\widehat{h}%
_{1},\;\;i=1,...,n,  \label{24}
\end{equation}%
where $\widehat{}$ \ denotes the restriction to $S$ and $h_{n+1}=0.$ Notice
that the $\rho _{i}(\lambda )$ are the coefficients of the minimal
polynomial of the Nijenhuis tensor:
\begin{equation}
(\det (\lambda I-N))^{1/2}=\sum_{i=0}^{n}\rho _{i}\lambda
^{n-i}=\prod_{i=1}^{n}(\lambda -\lambda _{i}),\;\;\;\rho _{0}=1.  \label{26a}
\end{equation}

There exists a sequence of separable ''potentials'' $V_{k}^{(r)},r=\pm 1,\pm
2,...,$ which can be added to $h_{k}(\lambda ,\mu ,c),$ given by the
following recursion relation \cite{5}
\begin{equation}
V_{k}^{(r+1)}=V_{k+1}^{(r)}-V_{k}^{(1)}V_{1}^{(r)},\;\;\;\;V_{k}^{(1)}=\rho
_{k},\;\;\;r=1,2,...,  \label{28}
\end{equation}%
and its inverse
\begin{equation}
V_{k}^{(-r-1)}=V_{k-1}^{(-r)}-V_{k}^{(-1)}V_{n}^{(-r)},\;\;\;\;V_{k}^{(-1)}=%
\rho _{k-1}/\rho _{n},\;\;\;r=1,2,...\;.  \label{28a}
\end{equation}%
Notice that recursion formulae are coordinate free and generate separable
potentials starting from the coefficients of the minimal polynomial of the
Nijenhuis tensor in arbitrary set of coordinates. Potentials $V^{(r)}$ (\ref%
{28}) and $V^{(-s)}$ (\ref{28a}) entrance the separation curve in
the following way
\begin{equation}
f(\lambda ,\mu )=\lambda ^{n+r-1}+c\lambda ^{n}+h_{1}\lambda ^{n-1}+\cdots
+h_{n}+\lambda ^{-s}.  \label{28b}
\end{equation}

\section{Canonical coordinates}

Now, let us consider an arbitrary canonical transformation on $M$
\begin{equation}
(\lambda ,\mu )\rightarrow (q,p)  \label{29}
\end{equation}%
independent of a Casimir coordinate $c$ (not necessarily a point
transformation!). The advantage of staying inside such a class of
transformations is that the clear structure of the pencil is preserved.

Applying the transformation (\ref{29}) to Hamiltonian functions (\ref{18})
and Poisson matrices (\ref{20}) one finds that
\begin{equation}
h_{k}(q,p,c)=h_{k}(q,p)+cb_{k}(q,p),\,\,\,k=1,...,n  \label{30}
\end{equation}%
and
\begin{eqnarray}
\pi _{0} &=&\left(
\begin{array}{cc}
\theta _{0} & 0 \\
0 & 0%
\end{array}%
\right) ,\,\,\,\theta _{0}=\left(
\begin{array}{cc}
0 & I \\
-I & 0%
\end{array}%
\right) ,  \nonumber \\
&&  \label{31} \\
\pi _{1} &=&\left(
\begin{array}{cc}
\theta _{1} & K_{1} \\
-K_{1}^{T} & 0%
\end{array}%
\right) ,\,\,\,\theta _{1}=\left(
\begin{array}{cc}
D(q,p) & A(q,p) \\
-A^{T}(q,p) & B(q,p)%
\end{array}%
\right) ,  \nonumber
\end{eqnarray}%
where $A,B$ and $D$ are $n\times n$ matrices. The nondegenerate Poisson
pencil $\theta _{\lambda }$ on $S$ gives rise to the related Nijenhuis
tensor $N$ and its adjoint $N^{\ast }$ in $(q,p)$ coordinates in the form
\begin{equation}
N=\theta _{1}\circ \theta _{0}^{-1}=\left(
\begin{array}{cc}
A & -D \\
B & A^{T}%
\end{array}%
\right) ,\,\,\,N^{\ast }=\theta _{0}^{-1}\circ \theta _{1}=\left(
\begin{array}{cc}
A^{T} & -B \\
D & A%
\end{array}%
\right) .  \label{33}
\end{equation}

Obviously, in a real situation we start from a given bi-Hamiltonian chain (%
\ref{30})-(\ref{33}) in canonical coordinates $(q,p,c),$ derived by some
method, and we try to find the DN coordinates which diagonalize the
appropriate Nijenhuis tensor and are separated coordinates for the system
considered. So now we pass to a systematic derivation of the inverse of
transformation (\ref{29}).

The first part of the transformation is given by
\begin{equation}
\rho _{i}(\lambda )=b_{i}(q,p),\;\;\;i=1,...,n.  \label{34}
\end{equation}%
The second part can be found in a few ways. One method was presented in
refs. \cite{7} and \cite{m1}. Here we present another method suggested in %
\cite{m2}. Consider the vector field $Y=\pi _{0}\circ db_{1}(q,p).$ In DN
coordinates it has the following form
\begin{equation}
Y=\sum_{i=1}^{n}\frac{\partial }{\partial \mu _{i}}=\pi _{0}\circ d\rho
_{1}(\lambda ),\,\;\;\;\rho _{1}=-\sum_{i=1}^{n}\lambda _{i}.  \label{35}
\end{equation}%
Hence, if some $\varphi _{i}$ depends only on a pair of coordinates $%
(\lambda _{i},\mu _{i}),$ then $Y(\varphi _{i})$ also depends only on $%
(\lambda _{i},\mu _{i})$. So, when for some $\overline{\varphi }_{i}$ we
have $Y(\overline{\varphi }_{i})=1,$ then according to the gauge (\ref{19})
it means that $\mu _{i}=\overline{\varphi }_{i}$ are an admissible DN
momenta and the second part of transformation is given by
\begin{equation}
\mu _{i}=\overline{\varphi }_{i}(q,p),\;\;\;i=1,...,n.  \label{36}
\end{equation}%
For this procedure our two basic objects, written in $(q,p)$ coordinates,
are
\begin{equation}
Y=\pi _{0}\circ db_{1}(q,p),\;\;\;\varphi _{i}=h_{\lambda _{i}}(q,p).
\label{37}
\end{equation}%
More details will be given in examples.

In a special case of a point transformation, when $b_{i}=b_{i}(q),%
\;i=1,...,n,$ the first part of the transformation is of the form
\begin{equation}
\rho _{i}(\lambda )=b_{i}(q)\Longrightarrow q_{i}=\alpha _{i}(\lambda
),\;\;\;i=1,...,n  \label{38}
\end{equation}%
and the second part can be constructed from a generating function $%
G(p,\lambda )=\sum_{i}p_{i}\alpha _{i}(\lambda )$ in the following way%
\begin{equation}
\mu _{i}=\frac{\partial G}{\partial \lambda _{i}}\Longrightarrow p_{i}=\beta
_{i}(\lambda ,\mu ),\;\;\;i=1,...,n.  \label{39}
\end{equation}

\section{Noncanonical coordinates: a general case}

When the Poisson chain (\ref{3}) is given in an arbitrary coordinate system $%
\{g_{i}\}_{i=1}^{2n+1}$ a clear structure of a pencil is lost and it is far
from obvious whether it is projectable onto a symplectic leaf $S$ of the
first Poisson structure or not. On the other hand, such a projectibility is
a necessary condition for separability of the chain, as DN coordinates are
these which diagonalize an appropriate Nijenhuis tensor on $S$ constructed
from a nondegenerated Poisson pencil on $S$. Here we adopt a method proposed
in \cite{m1}, \cite{m2} to sketch the simplest case of one-Casimir Poisson
pencils.

Let a vector field $Z$ be transversal to the symplectic foliation $S$ of $%
\pi _{0}.$ Consider the class of functions $\mathcal{F}(M)$ such that
\begin{equation}
\mathcal{L}_{Z}F=Z(F)=0,\;\;\;\forall F\in \mathcal{F}(M),  \label{40}
\end{equation}%
where $\mathcal{L}$ means a Lie derivative. We can identify $\mathcal{F}$
with all functions on some leaf $S_{0},$ as for an arbitrary $f\in \mathcal{F%
}(S_{0})$ one can define its extension $F$ on $M$ such that (\ref{40}) is
fulfilled. Hence
\begin{equation}
\mathcal{F}(S_{0})\ni f=F_{\mid S_{0}}.  \label{41}
\end{equation}%
We are looking for the condition on $\pi _{\lambda }$ such that
\begin{equation}
\forall F,G\in A\;:\;\;\{F,G\}_{\pi _{\lambda }}\in A.  \label{42}
\end{equation}%
Then, $\theta _{\lambda }$ defined as
\begin{equation}
\{f,g\}_{\theta _{\lambda }}:=\{F,G\}_{\pi _{\lambda }\mid S_{0}}  \label{43}
\end{equation}%
is a projection of $\pi _{\lambda }$ along $Z$ on $S_{0}.$

\textbf{Theorem 2} \emph{A sufficient condition for the projectability of $%
\pi _{\lambda }$ onto $S_{0}$ is
\begin{equation}
\mathcal{L}_{Z}\pi _{0}=0,\;\;\;\mathcal{L}_{Z}\pi _{1}=Y\wedge Z,
\label{44}
\end{equation}%
where $Y$ is some vector field.}

\textbf{Proof} Let $\pi $ be a Poisson tensor. Then $\forall F,G\in A:$
\begin{eqnarray}
\mathcal{L}_{Z}\{F,G\}_{\pi } &=&\mathcal{L}_{Z}<dG,\pi \circ dF>  \nonumber
\\
&=&<(\mathcal{L}_{Z}dG),\pi \circ dF>+<dG,(\mathcal{L}_{Z}\pi )\circ dF+\pi
\circ (\mathcal{L}_{Z}dF)>  \nonumber \\
&=&<d(\mathcal{L}_{Z}G),\pi \circ dF>+<dG,(\mathcal{L}_{Z}\pi )\circ dF+\pi
\circ d(\mathcal{L}_{Z}F)>  \nonumber \\
&=&<dG,(\mathcal{L}_{Z}\pi )\circ dF>.  \label{45a}
\end{eqnarray}%
For $\pi =\pi _{0}$ under the condition (\ref{44}) we have immediately $%
\mathcal{L}_{Z}\{F,G\}_{\pi _{0}}=0.$ For $\pi =\pi _{1}$ the condition (\ref%
{44}) gives
\begin{eqnarray}
\mathcal{L}_{Z}\{F,G\}_{\pi _{1}} &=&<dG,(\mathcal{L}_{Z}\pi _{1})\circ
dF>=<dG,(Y\wedge Z)\circ dF>  \nonumber \\
&=&<dG,(Y\otimes Z-Z\otimes Y)\circ dF>  \nonumber \\
&=&(Y\otimes Z\circ dF)G-(Z\otimes Y\circ dF)G  \label{46} \\
&=&Y(G)\cdot Z(F)-Z(G)\cdot Y(F)=0,  \nonumber
\end{eqnarray}%
so $\mathcal{L}_{Z}\{F,G\}_{\pi _{\lambda }}=0$ and the relation (\ref{42})
is fulfilled.

Moreover, the following theorem can be proved.

\textbf{Theorem 3} \emph{Let a Poisson pencil be projectible in the sense of
Theorem 2. If additionally
\begin{equation}
\mathcal{L}_{Z}(\mathcal{L}_{Z}h_{\lambda })=Z(Z(h_{\lambda }))=0  \label{47}
\end{equation}%
and vector fields $Z$ and $Y$ are normalized in such a way that
\begin{equation}
\mathcal{L}_{Z}h_{0}=Z(h_{0})=1,\;\;\;\mathcal{L}_{Y}h_{0}=Y(h_{0})=0,
\label{48}
\end{equation}%
then }

\begin{itemize}
\item[(i)] \emph{$h_{\lambda }$ is linear in a Casimir of $h_{0}$, }

\item[(ii)] \emph{$Y=\pi _{0}\circ d(Z(h_{1})),$ }

\item[(iii)] \emph{on $S_{0}$ the chain (\ref{3}) takes the form
\begin{equation}
N^{\ast }\circ \widehat{dh}_{i}=\widehat{dh}_{i+1}-Z(h_{i})\widehat{dh}%
_{1},\;\;i=1,...,n.  \label{49}
\end{equation}
}
\end{itemize}

Notice that this is a separating case from previous Sections, where now
\begin{equation}
\rho _{i}(\lambda )=Z(h_{i})  \label{50}
\end{equation}%
is the first part of a transformation to the DN coordinates on $S_{0}$. In
the language of the present Section, for arbitrary canonical coordinates
considered in the previous Sections, we have
\[
Z=\frac{\partial }{\partial c},\;\mathcal{L}_{Z}\pi _{0}=0,\;\;\mathcal{L}%
_{Z}\pi _{1}=Y\wedge Z,\;\;Y=\pi _{0}\circ d(Z(h_{1}))=\pi _{0}\circ d\frac{%
\partial h_{1}}{\partial c},\;
\]%
\begin{equation}
Z(Z(h_{\lambda }))=0,\;\;Z(h_{0})=1,\;\;Y(h_{0})=0.  \label{51}
\end{equation}

\section{Stationary flow of Harry-Dym}

Here we consider the following Newton equations of motion
\begin{equation}
q_{1xx}=8q_{1}^{-5}q_{2}+\alpha q_{1},\;\;\;q_{2xx}=-2q_{1}^{-4}+4\alpha
q_{2}-c,\;\;\;\;\;\alpha =const,  \label{52}
\end{equation}%
with $x$ as an evolution parameter, which are the second stationary flow of
the Harry-Dym hierarchy \cite{bmr}, \cite{3}. The appropriate bi-Hamiltonian
chain is the following
\begin{eqnarray*}
h_{0} &=&c, \\
h_{1} &=&\frac{1}{2}p_{1}^{2}+\frac{1}{2}p_{2}^{2}+2q_{1}^{-4}q_{2}-\frac{1}{%
2}\alpha q_{1}^{2}-2\alpha q_{2}^{2}+q_{2}c, \\
h_{2} &=&\frac{1}{2}q_{2}p_{1}^{2}-\frac{1}{2}q_{1}p_{1}p_{2}+\frac{1}{2}%
q_{1}^{-2}+2q_{1}^{-4}q_{2}^{2}+\frac{1}{2}\alpha q_{1}^{2}q_{2}-\frac{1}{4}%
q_{1}^{2}c,
\end{eqnarray*}%
\begin{eqnarray}
\pi _{0} &=&\left(
\begin{array}{ccc}
0 & I & 0 \\
-I & 0 & 0 \\
0 & 0 & 0%
\end{array}%
\right) ,\;  \label{53b} \\
\pi _{1} &=&\left(
\begin{array}{ccccc}
0 & 0 & 0 & -\frac{1}{2}q_{1} & p_{1} \\
0 & 0 & -\frac{1}{2}q_{1} & -q_{2} & p_{2} \\
0 & \frac{1}{2}q_{1} & 0 & \frac{1}{2}p_{1} & -8q_{1}^{-5}q_{2}-\alpha q_{1}
\\
\frac{1}{2}q_{1} & q_{2} & -\frac{1}{2}p_{1} & 0 & 2q_{1}^{-4}-4\alpha
q_{2}+c \\
-p_{1} & -p_{2} & 8q_{1}^{-5}q_{2}+\alpha q_{1} & -2q_{1}^{-4}+4\alpha
q_{2}-c & 0%
\end{array}%
\right) ,  \nonumber
\end{eqnarray}%
where $p_{1}=q_{1x},p_{2}=q_{2x}.$ This is the case of canonical coordinates
of Section 4 and the first part of the transformation (\ref{34}) to DN
coordinates is
\[
\rho _{1}=-\lambda _{1}-\lambda _{2}=q_{2},\;\;\;\rho _{2}=\lambda
_{1}\lambda _{2}=-\frac{1}{4}q_{1}^{2}
\]%
\begin{equation}
\;\;\;\;\;\;\;\;\;\;\;\;\;\;\;\;\;\;\Downarrow   \label{54}
\end{equation}%
\[
q_{1}=2\sqrt{-\lambda _{1}\lambda _{2}},\;\;q_{2}=-\lambda _{1}-\lambda _{2}.
\]%
Evidently this is a point transformation, so the second part of the
transformation can be constructed either through a generating function (\ref%
{39}) or by a general approach presented in Section 3. Because the first
method is standard we apply here the second one. As $Y=\pi _{0}\circ
d(q_{2})=-\frac{\partial }{\partial p_{2}}$, we have
\begin{equation}
Y(h_{\lambda })=-p_{2}\lambda +\frac{1}{2}q_{1}p_{1},\;Y(Y(h_{\lambda
}))=Y^{2}(h_{\lambda })=\lambda \;\Longrightarrow \;Y\left( \frac{%
Y(h_{\lambda })}{Y^{2}(h_{\lambda })}\right) =1.  \label{55}
\end{equation}%
It means that
\begin{equation}
\mu _{1}=\frac{Y(h_{\lambda _{1}})}{Y^{2}(h_{\lambda _{1}})}=-p_{2}+\frac{1}{%
2}q_{1}p_{1}\frac{1}{\lambda _{1}},\;\mu _{2}=\frac{Y(h_{\lambda _{2}})}{%
Y^{2}(h_{\lambda _{2}})}=-p_{2}+\frac{1}{2}q_{1}p_{1}\frac{1}{\lambda _{2}}
\label{56}
\end{equation}%
and hence
\begin{equation}
p_{1}=\sqrt{-\lambda _{1}\lambda _{2}}\left( \frac{\mu _{1}}{\Delta _{1}}+%
\frac{\mu _{2}}{\Delta _{2}}\right) ,\;p_{2}=-\lambda _{1}\frac{\mu _{1}}{%
\Delta _{1}}-\lambda _{2}\frac{\mu _{2}}{\Delta _{2}},\;\;\;\Delta
_{1}=-\Delta _{2}=\lambda _{1}-\lambda _{2},  \label{57}
\end{equation}%
\[
f(\lambda _{i},\mu _{i})=\frac{1}{2}\lambda _{i}\mu _{i}^{2}-2\alpha \lambda
_{i}^{3}+\frac{1}{8}\lambda _{i}^{-2}.
\]%
The separation curve takes the form
\begin{equation}
\frac{1}{2}\lambda \mu ^{2}-2\alpha \lambda ^{3}+\frac{1}{8}\lambda
^{-2}=c\lambda ^{2}+h_{1}\lambda +h_{2}.  \label{58}
\end{equation}

Let us now relate the presented approach to the Sklyanin one. It is known
that Liouville integrable systems can be put into the Lax form \cite{bab}
\begin{equation}
L_{x}+[L,U]=0,  \label{59}
\end{equation}%
where $L,U$ are some matrices, $[.,.]$ means the commutator and $x$ is an
evolution parameter. In the simplest case, when $L$ is $2\times 2$ traceless
matrix
\begin{equation}
L=\left(
\begin{array}{cc}
A(\lambda ;q,p) & B(\lambda ;q,p) \\
C(\lambda ;q,p) & -A(\lambda ;q,p)%
\end{array}%
\right) ,  \label{60}
\end{equation}%
i.e. in the case of the so-called Mumford systems \cite{mum}, $\lambda
_{i},i=1,...,n$ are roots of $B=0$ and $\mu _{i}=-A(\lambda
_{i};q,p),i=1,...,n.$ The separated coordinates are different points of the
\emph{spectral curve }$\det (L-\mu I)=0.$

A Lax pair for the stationary Harry Dym was found in \cite{bmr} in the form
\begin{eqnarray}
L &=&\left(
\begin{array}{cc}
-q_{1}p_{1}\lambda ^{2}+2p_{2}\lambda  & q_{1}^{2}\lambda ^{2}-4q_{2}\lambda
-4 \\
-q_{1}^{-2}\lambda ^{3}-(4q_{1}^{-4}q_{2}+p_{1}^{2})\lambda ^{2}+(4\alpha
q_{2}-2c)\lambda -4\alpha  & q_{1}p_{1}\lambda ^{2}-2p_{2}\lambda
\end{array}%
\right) ,  \nonumber \\
U &=&\left(
\begin{array}{cc}
0 & 1 \\
-4q_{1}^{-4}\lambda +\alpha  & 0%
\end{array}%
\right) .  \label{61a}
\end{eqnarray}%
On the other hand, it is well known that the Lax representation is not
unique and some admissible representation can be obtained for example via
the transformation: $\lambda \rightarrow \lambda ^{-1},L(\lambda
^{-1})\rightarrow \frac{1}{2}\lambda L(\lambda ^{-1}):$
\begin{eqnarray}
L &=&\frac{1}{2}\left(
\begin{array}{cc}
-q_{1}p_{1}\lambda ^{-1}+2p_{2} & q_{1}^{2}\lambda ^{-1}-4q_{2}-4\lambda  \\
-q_{1}^{-2}\lambda ^{-2}-(4q_{1}^{-4}q_{2}+p_{1}^{2})\lambda ^{-1}+8\alpha
q_{2}-2c-4\alpha \lambda  & q_{1}p_{1}\lambda ^{-1}-2p_{2}%
\end{array}%
\right) ,  \nonumber \\
U &=&\left(
\begin{array}{cc}
0 & 1 \\
-4q_{1}^{-4}\lambda ^{-1}+\alpha  & 0%
\end{array}%
\right) .  \label{62a}
\end{eqnarray}%
For this Lax representation the roots of $B(q,p;\lambda )=\frac{1}{2}%
q_{1}^{2}\lambda ^{-1}-2q_{2}-2\lambda =0$ and $\mu _{i}=-A(q,p;\lambda
_{i})=\frac{1}{2}q_{1}p_{1}\lambda ^{-1}-p_{2}$ are just the same separated
coordinates (\ref{54}), (\ref{56}) as in the bi-Hamiltonian approach and
moreover
\begin{equation}
\det (L-\mu I)=0\;\Longleftrightarrow \;\frac{1}{2}\lambda \mu ^{2}-2\alpha
\lambda ^{3}+\frac{1}{8}\lambda ^{-2}=c\lambda ^{2}+h_{1}\lambda +h_{2}.
\label{63}
\end{equation}%
Hence, in the case of stationary Harry Dym,
\begin{equation}
separation\;curve\;=\;spectral\;curve.  \label{64}
\end{equation}

\section{The KdV dressing chain}

Consider the so-called dressing chain
\begin{equation}
(v_{n}+v_{n+1})_{x}=v_{n}^{2}-v_{n+1}^{2}+\alpha _{n}  \label{65}
\end{equation}%
for the Schr\"{o}dinger equation
\begin{equation}
\Psi _{xx}=(u-\lambda )\Psi .  \label{66}
\end{equation}%
Actually, if $u_{n}$ is a sequence of solutions of (\ref{66}), generated by
a chain of Darboux transformations, and  $u_{n}=v_{nx}+v_{n}^{2}+\beta
_{n},\;$is a Miura map to the modified fields $v_{n}$, then, as was shown in %
\cite{sh}, the new fields $v_{n}$ are related among themselves through the
chain of equations (\ref{65}), where $\alpha _{n}=\beta _{n}-\beta _{n+1}.$
Let us close the chain (\ref{65})
\begin{equation}
v_{i}\equiv v_{i+N},\;\;\alpha _{i}\equiv \alpha _{i+N},  \label{67}
\end{equation}%
and assume that $\sum_{i=1}^{N}\alpha _{i}=0,$ then we obtain a finite
dimensional dynamical system
\begin{equation}
(v_{i}+v_{i+1})_{x}=v_{i}^{2}-v_{i+1}^{2}+\beta _{i}-\beta
_{i+1},\;\;\;i=1,...,N.  \label{68}
\end{equation}%
As was shown by Veselov and Shabat \cite{vs}, for $N=2n+1,$ it is a
bi-Hamiltonian system for which the bi-Hamiltonian chain (\ref{3}) can be
constructed. In $g_{i}=v_{i}+v_{i+1}$ coordinates the nonzero matrix
elements of both Poisson tensors are the following
\begin{eqnarray}
\{g_{i},g_{i-1}\}_{\pi _{0}} &=&1,  \nonumber \\
\{g_{i},g_{j}\}_{\pi _{1}} &=&(-1)^{j-i}g_{i}g_{j},\;\;\;j\neq i\pm 1,
\label{69} \\
\{g_{i},g_{i-1}\}_{\pi _{1}} &=&g_{i}g_{i-1}+\beta _{i},  \nonumber
\end{eqnarray}%
and the Casimir of the pencil $\pi _{1}-\lambda \pi _{0}$ is given by
\begin{eqnarray}
h_{\lambda } &=&h_{0}\lambda ^{n}+h_{1}\lambda ^{n-1}+...+h_{n}  \nonumber \\
&=&(-1)^{N}\left[ \prod_{j=1}^{N}\left( 1+\zeta _{j+1}\frac{\partial ^{2}}{%
\partial g_{i}\partial g_{j}}\right) \right] \prod_{k=1}^{N}g_{k},\;\;\;\;\;%
\;\zeta _{i}=\beta _{i}-\lambda .\;\;\;  \label{70b}
\end{eqnarray}

As $g_{i},i=1,...,N,$ are noncanonical coordinates, to separate the system
we have to apply the general formalism of Section 5. For $Z=\frac{\partial }{%
\partial g_{N}}$ one finds
\[
\mathcal{L}_{Z}\pi _{0}=0,\;\;\;\mathcal{L}_{Z}\pi _{1}=Z\wedge Y,
\]%
\begin{equation}
Y=\sum_{i=1}^{N-1}(-1)^{i+1}g_{i}\frac{\partial }{\partial g_{i}}+\left(
\sum_{i=1}^{N-1}(-1)^{i}g_{i}\right) \frac{\partial }{\partial g_{N}}=\pi
_{0}\circ d(Z(h_{1}))  \label{71}
\end{equation}%
\[
Z(Z(h_{\lambda }))=0,\;\;\;Z(h_{0})=1,\;\;\;Y(h_{0})=0.
\]%
The first part of the transformation to the DN coordinates is
\begin{equation}
c=h_{0},\;\;\;\rho _{k}(\lambda )=\frac{\partial h_{k}}{\partial g_{N}}%
,\;\;\;k=1,...,n.  \label{72}
\end{equation}%
The following property of the Casimir  $h_{\lambda }$
\begin{equation}
Y^{3}(h_{\lambda })=Y(h_{\lambda })\;\Longrightarrow \;Y(\ln (Y(h_{\lambda
})+Y^{2}(h_{\lambda }))=1  \label{73}
\end{equation}%
gives the second part of the transformation
\begin{equation}
\mu _{i}=\ln (Y(h_{\lambda _{i}})+Y^{2}(h_{\lambda _{i}})),\;\;\;i=1,...,n.
\label{74}
\end{equation}

Let us illustrate the method in the case  $N=5.$ We have
\begin{eqnarray}
h_{0} &=&g_{1}+g_{2}+g_{3}+g_{4}+g_{5},  \nonumber \\
h_{1}
&=&-g_{1}g_{2}g_{3}-g_{2}g_{3}g_{4}-g_{3}g_{4}g_{5}-g_{4}g_{5}g_{1}-g_{51}g_{1}g_{2}-g_{1}(\beta _{3}+\beta _{5})
\nonumber \\
&&-g_{2}(\beta _{4}+\beta _{1})-g_{3}(\beta _{5}+\beta _{2})-g_{4}(\beta
_{1}+\beta _{3})-g_{5}(\beta _{2}+\beta _{4}),  \label{75} \\
h_{2} &=&g_{1}g_{2}g_{3}g_{4}g_{5}+\beta _{1}g_{2}g_{3}g_{4}+\beta
_{2}g_{3}g_{4}g_{5}+\beta _{3}g_{4}g_{5}g_{1}+\beta _{4}g_{5}g_{1}g_{2}
\nonumber \\
&&+\beta _{5}g_{1}g_{2}g_{3}+\beta _{3}\beta _{5}g_{1}+\beta _{1}\beta
_{4}g_{2}+\beta _{2}\beta _{5}g_{3}+\beta _{1}\beta _{3}g_{4}+\beta
_{2}\beta _{4}g_{5},  \nonumber
\end{eqnarray}%
\begin{eqnarray}
\pi _{0} &=&\left(
\begin{array}{ccccc}
0 & -1 & 0 & 0 & 1 \\
1 & 0 & -1 & 0 & 0 \\
0 & 1 & 0 & -1 & 0 \\
0 & 0 & 1 & 0 & -1 \\
-1 & 0 & 0 & 1 & 0%
\end{array}%
\right) ,  \label{76a} \\
\pi _{1} &=&\left(
\begin{array}{ccccc}
0 & -g_{1}g_{2}-\beta _{2} & g_{1}g_{3} & -g_{1}g_{4} & g_{1}g_{5}+\beta _{1}
\\
g_{2}g_{1}+\beta _{2} & 0 & -g_{2}g_{3}-\beta _{3} & g_{2}g_{4} & -g_{2}g_{5}
\\
-g_{3}g_{1} & g_{3}g_{2}+\beta _{3} & 0 & -g_{3}g_{4}-\beta _{4} & g_{3}g_{5}
\\
g_{4}g_{1} & -g_{4}g_{2} & g_{4}g_{3}+\beta _{4} & 0 & -g_{4}g_{5}-\beta _{5}
\\
-g_{5}g_{1}-\beta _{1} & g_{5}g_{2} & -g_{5}g_{3} & g_{5}g_{4}+\beta _{5} & 0%
\end{array}%
\right) .  \nonumber
\end{eqnarray}%
Tthe transformation to DN coordinates is
\begin{eqnarray}
g_{1} &=&\frac{1}{2}\frac{(\lambda _{2}-\lambda _{1})e^{\mu _{1}+\mu _{2}}}{%
(\lambda _{2}-\beta _{3})(\lambda _{2}-\beta _{4})(\lambda _{2}-\beta
_{5})e^{\mu _{1}}-(\lambda _{1}-\beta _{3})(\lambda _{1}-\beta _{4})(\lambda
_{1}-\beta _{5})e^{\mu _{2}}},  \nonumber \\
g_{2} &=&2\frac{[(\lambda _{2}-\beta _{2})(\lambda _{2}-\beta _{4})(\lambda
_{2}-\beta _{5})e^{\mu _{1}}-(\lambda _{1}-\beta _{2})(\lambda _{1}-\beta
_{4})(\lambda _{1}-\beta _{5})e^{\mu _{2}}]}{(\lambda _{2}-\lambda
_{1})[(\lambda _{2}-\beta _{4})(\lambda _{2}-\beta _{5})e^{\mu
_{1}}-(\lambda _{1}-\beta _{4})(\lambda _{1}-\beta _{5})e^{\mu _{2}}]e^{\mu
_{1}+\mu _{2}}}  \nonumber \\
&&\times \lbrack (\lambda _{2}-\beta _{3})(\lambda _{2}-\beta _{4})(\lambda
_{2}-\beta _{5})e^{\mu _{1}}-(\lambda _{1}-\beta _{3})(\lambda _{1}-\beta
_{4})(\lambda _{1}-\beta _{5})e^{\mu _{2}}]  \nonumber \\
g_{3} &=&-\frac{1}{2}\frac{[(\lambda _{2}-\beta _{3})(\lambda _{2}-\beta
_{5})e^{\mu _{1}}-(\lambda _{1}-\beta _{3})(\lambda _{1}-\beta _{5})e^{\mu
_{2}}]}{[(\lambda _{2}-\beta _{3})(\lambda _{2}-\beta _{4})(\lambda
_{2}-\beta _{5})e^{\mu _{1}}-(\lambda _{1}-\beta _{3})(\lambda _{1}-\beta
_{4})(\lambda _{1}-\beta _{5})e^{\mu _{2}}]}  \label{77a} \\
&&\times \frac{\lbrack (\lambda _{2}-\beta _{4})(\lambda _{2}-\beta
_{5})e^{\mu _{1}}-(\lambda _{1}-\beta _{4})(\lambda _{1}-\beta _{5})e^{\mu
_{2}}]}{(\lambda _{2}-\lambda _{1})(\lambda _{1}-\beta _{5})(\lambda
_{2}-\beta _{5})}  \nonumber \\
g_{4} &=&-2\frac{(\lambda _{2}-\lambda _{1})(\lambda _{1}-\beta
_{4})(\lambda _{1}-\beta _{5})(\lambda _{2}-\beta _{4})(\lambda _{2}-\beta
_{5})}{(\lambda _{2}-\beta _{4})(\lambda _{2}-\beta _{5})e^{\mu
_{1}}-(\lambda _{1}-\beta _{4})(\lambda _{1}-\beta _{5})e^{\mu _{2}}},
\nonumber \\
g_{5} &=&c-g_{1}-g_{2}-g_{3}-g_{4}.  \nonumber
\end{eqnarray}%
The appropriate function $f$ takes the form
\begin{equation}
f(\lambda _{i},\mu _{i})=2(\lambda _{i}-\beta _{1})...(\lambda _{i}-\beta
_{5})e^{-\mu _{i}}+\frac{1}{2}e^{\mu _{i}}  \label{78}
\end{equation}%
and the separation curve is
\begin{equation}
2(\lambda -\beta _{1})...(\lambda -\beta _{5})e^{-\mu }+\frac{1}{2}e^{\mu
}=c\lambda ^{2}+h_{1}\lambda +h_{2}.  \label{79}
\end{equation}

The case $N=5$ suggests the general form of a separation curve for an
arbitrary odd $N$
\begin{equation}
2(\lambda -\beta _{1})...(\lambda -\beta _{N})e^{-\mu }+\frac{1}{2}e^{\mu
}=c\lambda ^{n}+h_{1}\lambda ^{n-1}+...+h_{n}.  \label{80}
\end{equation}%
The implicit form of the trajectories $\lambda _{i}(t_{r}),$ calculated from
the solution of HJ equations, is
\begin{equation}
\sum_{k}\int^{\lambda _{k}}\frac{\xi _{i}^{n-i}}{\sqrt{h_{\xi
_{k}}^{2}-\gamma (\xi _{k})}}d\xi _{k}=\delta
_{ir}t_{r}+const_{i},\;\;\;i=1,...,n,  \label{81}
\end{equation}%
where
\begin{eqnarray*}
h_{\xi _{k}} &=&c\xi _{k}^{n}+a_{1}\xi _{k}^{n-1}+...+a_{n}, \\
\gamma (\xi _{k}) &=&4(\xi _{k}-\beta _{1})...(\xi _{k}-\beta _{N}).
\end{eqnarray*}

The Lax representation (\ref{59}) of the KdV dressing chain (\ref{68}) is
given by
\begin{equation}
L=L_{1}...L_{N},\;\;\;L_{i}=\left(
\begin{array}{cc}
v_{i} & 1 \\
v_{i}^{2}+\zeta _{i} & v_{i}%
\end{array}%
\right) ,  \label{82}
\end{equation}%
so, it is not a Mumford system as the $L$ matrix is not a traceless one. The
spectral curve takes the form
\begin{equation}
\det (L-\overline{\mu })=0\;\Longleftrightarrow \;\overline{\mu }^{2}-(trL)%
\overline{\mu }+\det L=0,  \label{83}
\end{equation}%
where
\begin{eqnarray}
\det L &=&(-1)^{N}\zeta _{1}...\zeta _{N}=(-1)^{N}(\lambda -\beta
_{1})...(\lambda -\beta _{N}),  \nonumber \\
trL &=&\tau _{N}=(-1)^{N}h_{\lambda }=(-1)^{N}(h_{0}\lambda
^{n}+h_{1}\lambda ^{n-1}+...+h_{n}).  \label{84}
\end{eqnarray}%
Obviously
\begin{equation}
separation\;curve\;\neq \;spectral\;curve  \label{85}
\end{equation}%
and points $\{\lambda _{i},\overline{\mu }_{i}\}_{i=1}^{n}$ from a spectral
curve are not canonical separated coordinates. Nevertheless, a simple
transformation of $\overline{\mu }$
\begin{equation}
(-1)^{N}\overline{\mu }=\frac{1}{2}e^{\mu }  \label{86}
\end{equation}%
transforms the spectral curve (\ref{83}) into the separation curve (\ref{79}%
), making the points $\{\lambda _{i},\mu _{i}\}_{i=1}^{n}$ canonical
separated coordinates.


\begin{thebibliography}{99}
\bibitem{sk1} E. K. Sklyanin, \emph{Separation of variables. New trends, }%
Prog. Theor. Phys. Suppl. \textbf{118 }(1995) 35

\bibitem{1} M. B\l aszak, \emph{On separability of bi-Hamiltonian chain with
degenerated Poisson structures,} J. Math. Phys. \textbf{39, }3213 (1998)

\bibitem{2} M. B\l aszak, \emph{Bi-Hamiltonian separable chains on
Riemannian manifolds, }Phys. Lett. A \textbf{243, }25 (1998)

\bibitem{3} M. B\l aszak, \emph{Multi-Hamiltonian Theory of Dynamical
Systems, } in: Texts and Monographs in Physics, Springer-Verlage (1998)

\bibitem{5} M. B\l aszak, \emph{Theory of separability of multi-Hamiltonian
chains, }J. Math. Phys. \textbf{40} (1999) 5725

\bibitem{4} M. B\l aszak, \emph{Inverse bi-Hamiltonian separable chains, }J.
Theor. Math. Phys. \textbf{122 }(2000) 140

\bibitem{6} M. B\l aszak, \emph{Separability of two-Casimir bi- and
tri-Hamiltonian chains,\thinspace } Rep. Math. Phys. \textbf{46 }(2000) 35

\bibitem{7} M. B\l aszak, \emph{Degenerate Poisson Pencils on Curves: New
Separability Theory,} J. Nonl. Math.Phys. \textbf{7 }(2000) 213

\bibitem{m1} G. Falqui, F. Magri and G. Tondo, \emph{Reduction of
bihamiltonian systems and separation of variables: an example from the
Boussinesq hierarchy, } Theor. Math. Phys. \textbf{122 }(2000) 176

\bibitem{m2} G. Falqui, F. Magri, M. Pedroni, \emph{Bihamiltonian geometry
and separation of variables for Toda lattices, }eprint \ nlin.SI/0002008
(2000)

\bibitem{gel} I. M. Gel'fand and I. Zakharevich, \emph{On the local geometry
of a bi-Hamiltonian structure. }In the Gelfand Mathematical Seminars
1990-1992 (L. Corwin et al. eds.), Birkauser, Boston, pp. 51-112 (1993)

\bibitem{bmr} S. Rauch-Wojciechowski, K. Marciniak and M. B\l aszak, \emph{%
Two Newton decompositions of stationary flows of KdV and Harry Dym
hierarchies, }Physica A \textbf{233 }(1996) 307

\bibitem{bab} O. Babelon and C.M. Viallet, \emph{Hamiltonian structures and
Lax equations, }Phys. Lett. \textbf{B} \textbf{237 }(1990) 411

\bibitem{mum} D. Mumford, \emph{Tata lectures on theta II, }Birkh\"{a}user
(1984)

\bibitem{sh} A. B. Shabat, \emph{The infinite-dimensional dressing dynamical
system, }Inverse Problems \textbf{6 }(1992) 303

\bibitem{vs} A. Veselov and A. B. Shabat, \emph{Dressing chain and spectral
theory of Schr\"{o}dinger operator, }Funk. analiz i pril. \textbf{27} \
(1993) 1.
\end{thebibliography}
\end{document}